# Principles and Framework for the Operationalisation of Meaningful Human Control over Autonomous Systems

Simeon C. Calvert*

* s.c.calvert@tudelft.nl;
Dept. of Transport & Planning; Delft University of Technology, The Netherlands
Stevinweg 1, 2628CN, Delft, The Netherlands. ORCID: 0000-0002-1173-0071

*Abstract* – With a plethora of different seemingly diverging expansions for use of Meaningful Human Control (MHC) in practice, this paper proposes an alignment for the operationalisation of MHC for autonomous systems by proposing operational principles for MHC and introducing a generic framework for its application. The increasing integration of autonomous systems in various domains emphasises a critical need to maintain human control to ensure responsible safety, accountability, and ethical operation of these systems. The concept of MHC offers an ideal concept for the design and evaluation of human control over autonomous systems, while considering human and technology capabilities. Through analysis of existing literature and investigation across various domains and related concepts, principles for the operationalisation of MHC are set out to provide tangible guidelines for researchers and practitioners aiming to implement MHC in their systems. The proposed framework dissects generic components of systems and their subsystems aligned with different agents, stakeholders and processes at different levels of proximity to an autonomous technology. The framework is domain-agnostic, emphasizing the universal applicability of the MHC principles irrespective of the technological context, paving the way for safer and more responsible autonomous systems.

Keywords: *Meaningful Human Control; Autonomous Systems; Human-Machine Interaction; Ethical AI*

# 1 Introduction

Autonomous systems interacting with humans and performing highly complex tasks are on the increase and are expected to increase in dominance in society. Maintaining a minimal level of control over such autonomous systems is vital to ensure safety and proper operation of these systems. Meaningful Human Control (MHC) is a concept that describes how humans can exert control over an autonomous system even when they are not in operational control. However, as a philosophical concept, MHC currently does not yet provide sufficient explicit guidelines how to be applied in practice for safety critical autonomous systems (Theodorou & Dignum 2020; Jensen 2020). Ekelhof (2019) states this eloquently that "abstract concepts are of little use if they ignore the operational context that confronts … their application". While the premise is that autonomous systems should adhere to MHC-like norms, which are necessary to uphold a balance of human moral responsibility, accountability, and ethical governance over autonomous systems, ensuring transparency, safety, and alignment with human values while preventing harm and legal responsibility gaps. This paper presents a generic framework for the operationalisation of MHC for autonomous systems based on derived principles for MHC operationalisation. The framework allows industry and scientific stakeholders alike to further detail the required conditions to the context of their systems' applications to ensure safe and humanly acceptable behaviour of autonomy and adhere to the three pillars of accountability, responsibility and transparency.

The concept of MHC originated in the political debate on autonomous weapons systems (Article 36 2013; Horowitz & Scharre 2015). It prescribes the conditions for a relationship between controlling human agents and a controlled autonomous system that preserves moral responsibility and clear human accountability even in the absence of any specific form of human operational. Santoni de Sio & Van den Hoven (2018) distinguish two key conditions for human control to be meaningful, namely the *tracing and tracking* conditions. These two conditions reflect 1) Tracing: the presence and role of one or more humans that are able to exert control over an autonomous system and harbour moral responsibility for the actions of the system, and 2) Tracking: the ability of the autonomous system to act responsibility and adhere to human reasons and intensions.

To date, the translation of MHC into a generalised approach for the operationalisation of MHC has not sufficiently been made, initially through a lack of understanding of the concept of MHC and how it connects to the physical and digital world. Santoni de Sio & Van den Hoven (2018) state that "policymakers and technical designers lack a detailed theory of what "meaningful human control" exactly means." Despite there being a consensus that autonomous systems should be under MHC (Ekelhof 2019), Horowitz & Scharre (2015) have previously been critical of the continued use of MHC concept while consensus and a clear tangible route to application is missing. Kwik (2022) also highlights that the international community appears keen to apply MHC, and that "crystalising MHC into a concrete framework is a paramount first step". In the meantime, various interpretations and derivations of MHC have appeared that in turn have led to an apparent divergence rather than convergence for application. For this reason, clear generic principles for MHC operationalisation are required (Horowitz & Scharre 2015). And while workable frameworks have been also proposed, primarily from Autonomous Weapons Systems (AWS) and Autonomous Driving Systems (ADS) domains (Calvert et al. 2024; Kwik 2022), they are too domain-specific to be easily applicable to other domains without further generalisation, but nevertheless, do give good initial directions and contain relevant elements that can be used as a basis to form a generic approach for MHC.

This paper goes further by posing the question how MHC can be effectively operationalised in autonomous systems? And thereby, the paper proposes an approach for the application of MHC to any autonomous system and aims to bring convergence and alignment in the concept of MHC in the form of principles for the operationalisation of MHC. Derivation of such principles acts as a stepping stone to formalise frameworks for how MHC can be used for design and evaluation purposes, finally resulting in guidelines for those wishing to apply MHC in practice. Conversely, the second main contribution lies in the construction of a generic framework for operationalisation of MHC for autonomous technology, based on the operational principles for MHC.

The approach taken in this research exists of four steps and makes use of elements from an approach often called conceptual synthesis[1] (Andreasen et al. 2015; Jabareen 2009; Nutley et al. 2002). Conceptual synthesis focusses on how concepts are generated, refined, and synthesised. While there are different ways to approach this, the approach followed in this paper emphasizes iterative refinement, decomposition of theoretical constructs, and systematic integration of concepts into an operational model (Andreasen et al. 2015). Nutley et al. (2002) also highlights that the intention of conceptual synthesis is not to provide an exhaustive search and review of all the literature published in a field, but rather the aim is to identify the key ideas, models and debates, and review the significance of these for developing a better understanding for implementation, which is reflected in steps 1 and 2. The first step (1) involves a structured literature review that identifies and analyses three relevant case domains. These domains are selected as safety-critical domains where autonomous systems are becoming more prevalent and scientific work with regard to MHC has started to develop. Thereafter in step (2), concepts related and aligned to MHC are reviewed to ensure breadth in the analysis. By tracing the theoretical evolution and comparing interpretations across various concepts, the analysis distinguishes MHC from related concepts and allows differences and points of agreement, and mutual strengths to be identified, as well as systematically addressing ambiguities, for example to prevent conflation. Using the insights from the related concepts, in step (3), a conceptual synthesis approach is applied to derive the key principles most relevant to the operationalisation of a broad definition of MHC and which are in agreement with literature, while also being actionable in an operational framework. The fourth step (4) is the construction of the generic framework based on the principles and best-practice from literature. This encompasses components of systems and their subsystems aligned with different agents, stakeholders and processes at different levels of proximity to an autonomous technology. Further detailing of each step is given at the start of each relevant section in the paper.

The paper starts by reviewing the developments of MHC in three distinct application domains in Section 2. In Section 3, the main related concepts to MHC are reviewed and followed in Section 4 by the deriving the main principles for MHC operationalisation. In Section 5, the framework for MHC operationalisation and its approach is presented, to conclude with the discussion in Section 6. Throughout. it should also be noted that the term ‘autonomous’ is used in this paper to define systems that can independently perform tasks without or with limited assistance of humans. In certain domains autonomous aligns with the term fully automated, while the terms semi-automation, partially automated or conditionally automated also exist to indicate an autonomous system is limited either to specific functionality or in symbiosis with a human agent. For the sake of consistency, we continue to use the term autonomous, while in some cases various other forms of automation may be closer to the common terminology used in certain domains.

[1] The same and similar approaches can also be found under other names in literature.

# 2 MHC application areas and concepts

The concept of MHC has been increasingly applied and considered in various complex socio-technical domains, increasingly going beyond its original beginnings in the defence domain. With this, new insights are gained of what MHC entails and how it can be applied in practice for different purposes and in different challenges. In this section, we consider how MHC has been applied and developed in three broadly defined domains with a specific focus on developments that are beneficial to operationalisation of MHC in practice. The choice of these domains was made from an initial shortlist of safety-critical seven domains where autonomous systems are becoming more prevalent and scientific work on MHC has started to develop. Only three domains are selected to give a demonstration of the practical challenges and the current work performed relating to MHC, without over-elaborating beyond the objective of this research. While other domains could have been included, such as energy or manufacturing, we found these three areas to be the most advanced with regard to the consideration of MHC in technical and autonomous systems and to be most representative to the current state of the art. To ensure a concise, but relevant review, the below descriptions touch upon some of the main and current activities relating to autonomous control and MHC, mentioning areas of application and the main theoretical and practice-orientated approaches towards operationalisation, aligning with the objective of this paper.

## *2.1 Defence*

The concept of MHC was first introduced in the military domain where a sense of urgency was present to act to setup constraints for 'autonomous systems of death'. Consensus has been reached on the most basic requirements of MHC: that an AWS must be "predictable, reliable and transparent technology, while providing accurate information for the user on the outcome sought, operation and function of technology, and the context of use" (Roff & Moyes 2016; Ekelhof 2019) and "timely human action and a potential for timely intervention, as well as accountability to a certain standard" is required (Roff & Moyes 2016). Nevertheless, concrete elaboration that allows MHC to be explicitly applied in practice still lacks amongst the discussions. In an attempt to address this, Kwik (2022) proposed an integrated framework as "a workable foundation for addressing many concerns related to the use of AWS". The framework revolves round two central interactive elements: the 'System' and the 'Operational Environment'. Various facets of AWS are identified and are connected to the primary human agent, the Operator, and to the AWS system. The approach acts as a basis for the further testing and refinement in practice, especially regarding legal aspect on accountability and responsibility. One element that does appear striking is the lack of additional human agents in the entire framework. The focus of the framework from a human control perspective is firmly on the Operator, while in practice many other human agents can influence the AWS in different proximal ways. Moreover, the distal influence not considered, which is believed to be a deliberate constraint by the authors, which entails aspects such as societal and governmental influence. Amoroso & Tamburrini (2020) propose an approach focussed on the alignment of MHC with International Human Law (IHL) that a human must be a *fail-safe actor*, *accountability attractor*, and *moral agency enactor*. These are applied to AWS with the control policies:

- Boxed autonomy: A human agent constrains the system to an operational box.
- Denied autonomy: All critical events are controlled by a 'fail-safe' human.
- Supervised autonomy: Humans monitor the AWS at all times and intervene when required.

Sharkey (2016) and adapted by Amoroso & Tamburrini (2020), proposes a taxonomy of increasing autonomy on a scale from full human control (L1) to full autonomous control (L5), with various intermediate combinations for target selection, engagement and initiation. Amoroso & Tamburrini (2021) later expand this to develop a framework that applies rules to ensure MHC is adhered to. They propose that rules are conceived as 'if-then' statements, where the 'if' statement includes 'what-where-how' properties connected to the context and operation of the automated system, such as "*what* mission the weapon system is involved with, *where* the system will be deployed, and *how* it will perform its tasks". The 'then' part connects the context and automation states to an appropriate human action for control. In such a way, Amoroso & Tamburrini (2021) connect the AWS to human actions and implicitly approach aspects of the tracking and the tracing conditions of MHC. Ekelhof (2019) takes a complementary, angle to boxed, denied and supervised autonomy, highlighting that distributed control is key in the discourse of AWS to maintain MHC, as the distributed nature of control illustrates that human control does not need to have a direct link with the weapon system. Ekelhof (2019) suggests that a process that recognizes the distributed nature of control in military decision-making is required. This again highlights the necessity to consider the whole chain of control, including those human agents that can exert control through decisions and actions that are less proximal to the operations of an autonomous system.

Extensive dialogue has been present relating MHC to AWS, however without a clear route to application according to many. Many of these discussions were initiated at the level of NGOs and international organisations campaigning for control over the automation and inclusion of AI in AWS (Borrie 2016; Crootof 2016; Horowitz & Scharre 2015) and MHC was quickly picked up by the community as a promising concept to connect autonomous control to human values (Boothby 2019; Crootof 2016; Ekelhof 2019; Gaeta 2016; Horowitz & Scharre 2015). Despite this, discussions and progress on its implementation in practice have been frustrated by a lack of progress (Jensen 2020; Schuller 2017)

## 2.2 Automated Driving

Possibly one of the areas that has seen the most MHC applied research outside of AWS is that of Automated Driving, as traffic is often a complex and human-critical environment for an automated system. Mecacci & de Sio (2019) state the "urgent practical issue" is that the human agent gives up a part of control to an autonomous vehicle, which has resulted in responsibility (Matthias 2004; Santoni de Sio & Mecacci 2021; Sparrow 2007) or accountability gaps (Heyns 2014), but maybe even more worryingly to lethally dangerous situations in with no clear human control (Calvert et al. 2020; Mecacci & de Sio 2019). Similarly to the taxonomy of increasing autonomy in AWS (Sharkey 2016), the Society of Automotive Engineers (SAE) (SAE 2018) have developed a globally accepted taxonomy of levels for automated driving. These levels describe the role of the autonomous system versus the human driver, with L0 being full human control, L1-2 are considered to be shared control, L3 supervised autonomy, and in L4-5 full operational control lies with the autonomous system with distal human monitoring at most.

In the domain of Automated Driving Systems (ADS), some significant steps have been made to operationalise MHC. Mecacci & de Sio (2019) describe MHC in ADS in terms of strategic, tactical to operational control (Michon 1985), which allows for an easier distinction to be made between different levels control agents through different types of mechanisms. Another key step saw the construction of the proximity scale, which describes human reasons mapped to specific human agents in alignment with the tracking condition (Mecacci & de Sio 2019). A distinction is made between *distal reasons*, which describe why a system may adopt a strategy, and *proximal reasons*, which describe how a system applies a strategy. For example, society has distal value and norms, while a driver of a vehicle has proximal reasons in their control of a vehicle. Calvert et al. (2024) extended the proximity scale to include the ADS (automated vehicle) and the surrounding environment and in doing so also demonstrated that this approach can be used to include aspects of the tracing condition (shown in Figure 1).

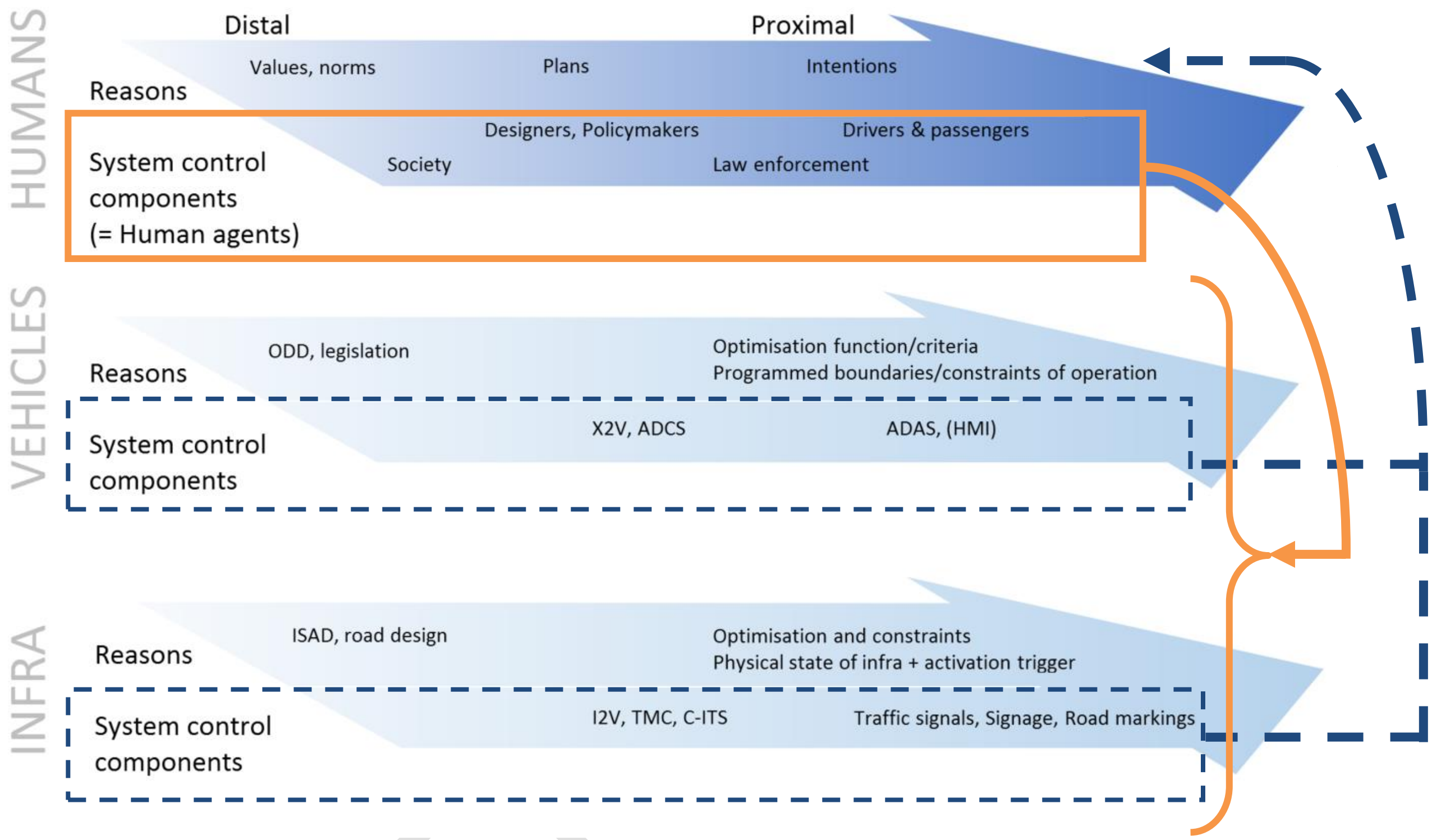

**Figure 1: Integrated system proximity framework for MHC over autonomous vehicles (adapted from Calvert et al. (2024)). Tracing is given in orange solid lines; Tracking is given in dark blue dashed lines.**

The proximity scale remains at an abstract philosophical-psychological level, which led Calvert et al. (2024) to prose an approach for the quantitative evaluation of the tracing condition. The approach focusses on the detailed identification of the various components of the autonomous driving systems, which includes the human driver, the vehicle, and the traffic environment. From this, the authors proposed a cascade model that evaluates the extent that each potential human agent involved with the autonomous system can exert meaningful control. The resulting score offers a tangible score for the presence of MHC over system. Calvert & Mecacci (2020) went further in the formulation of a comprehensive *taxonomy of tracking and tracing* conditions of MHC combined with the proximity scale and an explicit application to human reasons and behaviour. The authors demonstrated that the taxonomies form a solid and comprehensive foundation for further quantitative and qualitative operationalisation of MHC in engineering systems. Moreover, the taxonomy has broader general application for MHC beyond the context of ADS as breakthrough research compassing all the advancements made previously on the topic.

While the previous efforts focused on the mechanisms and evaluation of MHC, Calvert et al. (2024) proposed an Operational Process Design (OPD) approach aimed at generating greater understanding of how autonomous vehicle systems can be designed to incorporate a greater degree of MHC. The OPD approaches the problem from a systems approach in which different sub-systems are identified at which different levels of human control can be exerts, from distal through to proximal control. The OPD shows for Automated Driving how distal human agents, such as vehicle designers, regulators or society, can exert control through *explicit distal updating*. Proximally, both a human driver as well as the vehicle (through the automated driving control system) can be designed to also continuously improve the extent of MHC through *implicit proximal updating*. Driver experience and training is a key aspect that can improve control with an automated system, while if the automated vehicle operates using AI, it can be assumed that it is learning and improving its ability to perform better and adhere to human reasons to a better extent.

In this sub-section we have seen that beyond AWS, there are various areas in industrial engineering that have taken and advanced the concept of MHC beyond its initial beginnings as a philosophical concept. Especially recent developments in the past years in the area of Automated Driving have led to a greater understanding of how MHC can be applied for both evaluation and design of autonomous systems.

## 2.3 Healthcare

With the use of Robot-Assisted Surgery (RAS), the concept of MHC has started to emerge in the healthcare domain in recent years. Within the domain, it is recognised that areas like robotic surgery have not reached as an advanced level in robotic system autonomy compared to other domains (Ficuciello et al. 2019). For surgery, RAS will usually operate in a master slave control mode such that the behaviours of surgical robots emulate from a surgeon's hands-on supervision and real-time overriding authority. The RAS directly obeys the surgeon, hence the system is directly under the human control of the surgeon and also their responsibility and accountability. Ficuciello et al. (2019) states for this reason it is "unsurprising that the ethical discussion of surgical robot autonomy is still in its infancy and mostly embedded into technologically distant scenarios of highly autonomous systems." Further developments in microsurgery that robots can autonomously perform with sub-millimetre precision are on the horizon and that further benchmarking and policy is required. The next steps will involve "automating selected tasks using sensors and real-time feedback" to ensure human control (Ficuciello et al. 2019). Hierarchical levels of surgical robot autonomy are presented from no autonomy (L0) to robotic assistants that can constrain or correct human action (L1), robotic systems that carry out tasks designated by humans and under human supervision (L2), and robotic systems generate tasks execution strategies under human supervision (L3). A further L4 is defined as robots that autonomously perform an entire procedure under human supervision and L5 which requires no human supervision.

Beyond operational involvement of automation, the use of artificial intelligence-driven decision support systems (AI-DSS) are more prevalent in healthcare (Braun et al. 2021). These systems are used to provide tools to help clinicians as well as the patients to make better decisions in various processes, such as providing diagnoses (Castaneda et al. 2015), forecasts (Chen & Asch 2017) or treatment recommendations (Komorowski et al. 2018). The AI-DSS have the advantage that they often perform better or at least as good as physicians, especially for complex analysis, such as dermatology analysis (Gulshan et al. 2016; Haenssle et al. 2018) or radiology (Adams et al. 2021;

López-Cabrera et al. 2021). Braun et al. (2021) highlight and discuss the “entanglement” of AI-DSS with four normative notions relating to trustworthiness, transparency, agency and responsibility. Braun et al. (2021) argue that in the end AI-DSS are auxiliary tools to enhance human decision-making and that human agents should retain decisional authority, while recognising the benefits of using the systems.

It is interesting that in healthcare domain, operational control is less of an issue, but rather decision making plays a larger role. Which human agents are responsible and take responsibility, either through automated system design or as the knowledgeable user of the system is the more potent question, which focussed more closely on tracing condition of MHC. Up to this point, frameworks for the explicit application of MHC in healthcare have not yet been developed, as there is always a clear human controller in charge of a support system. As the level of automation advances, more scrutiny will be required to ensure MHC is present and maintained.

# 3 Related concepts and their applications

Starting this section, an immediate word of caution is given: Considering alternative concepts to MHC is a potentially endless search that can very quickly diverge into various domains, sub-domains and paths of thought from technology, automation, AI, psychology, human factors, philosophy, ethics and beyond. Therefore, the limited scope given here is a sub-set of the main concepts that have been considered in the past decade that closely align to the premise of MHC and control over autonomous system, which will allow us a greater alignment and focus for the practical application of MHC. There are different ways to make a distinction in concepts that focus on control over automation. We consider five related concepts to MHC that highlight various aspects of autonomous control from a perspective of human control and responsibility, and aid the process of developing a generic operationalisation framework for MHC. The considered concepts are Meaningful Human Certification (MHCrt), Responsibility and accountability, Comprehensive Human Oversight (CHO), Contestable AI by Design, and Value Centred Design (VCD).

The approach followed to come to the above selection of concepts involved literature review and expert consultation. Literature was reviewed relating explicitly to MHC as well as other concepts that consider control over autonomous systems from an ethical, behavioural and philosophical point of view. This resulted in a range of potential concepts. Thereafter, six different experts from ethics and philosophy of AI, as well as other experts related to discussions on autonomous systems were consulted. This consultation focussed on the suitability of the concepts found in literature, whether certain other concepts were missing and what the necessary scope would be for selection. Based on this, the focus is on concepts that consider autonomous control from a perspective of human control and responsibility and connect with MHC. Each of the selected concepts do this in different ways and hence contribute to the discussion. The further analysis in this section focuses on extracting a clear description of each concept, deriving the key elements at the heart of each concept and describing the relation to MHC such that the concepts can aid further development of MHC towards operationalisation. The overview of this result is shown in Table 2.

**Certification**

Meaningful Human Certification has achieved increased attention (Cummings 2019; Skeete 2021), especially for AWS, and emphasizes the need for rigorous training and certification of both autonomous systems and the individuals overseeing their deployment to ensure ethical, legal, and safe use. Skeete suggests a two-step certification for offensive AWS: a strategic decision by a high-level human, followed by the deployment of an autonomous system capable of outperforming humans in target engagement. However, the author notes that certifying autonomous systems for superior performance in safety-critical tasks remains unresolved in both military and civilian settings. Additionally, predicting their operational efficiency in dynamic environments remains a major challenge. Also on a strategic level, some form of MHCrt has the potential to increase accountability. A focus on both strategic and technological design certification is therefore crucial (Skeete 2021).

Cummings (2019) argues that in search of a better performance of human-machine interaction for AWS, MHC does not suffice due to a lack of control from humans. Humans can make mistakes when working with automation, while automated systems for various tasks outperform humans and should be certified to take on these tasks. Many autonomous systems are rarely fully controlled without human intervention, which means that on a design level, decisions need to be made where control should lie, which in turn should make up part of the certification process. While this may suffice on a practical level at present, it can be argued that certification is also part of the MHC concept aligned with the tracing condition as it attributes a human role to the certifier. The automated system still needs to adhere to human reasons and intentions, and humans still play an active role on a distal level, for example through certification, and on a proximal level through involvement with the AWS. The case put forward for MHCrt therefore appears to be based on the concept of MHC prior to the further elaborations that have since taken place with regard to operationalisation of the concept (e.g. (Calvert et al. 2019; Kwik 2022)). The latter developments therefore have extended the concept of MHC to also include MHCrt such that they are both in agreement rather than alternative approaches, at least within the realm of engineering. MHCrt does go further than MHC in the sense that certification gives a clear outline for legal and accountability. Steps for certification can be taken, while challenges still remain focussed on the ability to determine to what level a system should be certified (how well should it perform) and the absence of established methodologies by engineers to rigorously test these systems to identify and rectify both errors of commission and omission (Skeete 2021).

**Responsibility, accountability and Comprehensive Human Oversight**

With respect to responsibility and trustworthiness, Yazdanpanah et al. (2023) argue that to certify the legality of AI systems, concepts like responsibility, blame, accountability, and liability must be formalized and computationally implementable to address responsibility gaps. They stress the need for a balance in the design, development, and deployment of trustworthy autonomous systems (TAS) that allows for practical implementation, while being expressive enough to capture the sociotechnical nature of TAS. A key element to achieve this is ensuring that multiple agents can exert control either as latency or as shared control. On an operational level, this can be referred to as human–machine teams (Flemisch et al. 2016), or as a symbiotic in which autonomous system and human user co-control in association (Inga et al. 2023; Abbink et al. 2018). Ekelhof (2019) also highlights that there must be a matter of trust between operators and their superiors as well as the systems for effective operations, while realising that trusting the process or system in itself is not the same as exercising meaningful control. But also, on a strategic and design level, different teams and organisations will be involved in the design and ownership of a system, which demands a focus on joint co-creation and lines of responsibility. To that extent Yazdanpanah et al. (2023) describe the co-active design method, coined by Johnson et al. (2014), which includes the principles:

1. additional **monitoring** (to enhance mutual observability) functionalities,
2. agents **taking over** tasks from other team members (to improve resilience),
3. team members **informing and directing** other agents (to support mutual directability) based on insights in upcoming complications and
4. agents **knowing how** the collaborating agents work (to establish mutual predictability).

Verdiesen et al. (2021) take a different angle as they focus on accountability and define this as a form of responsibility. They propose a framework for Comprehensive Human Oversight (CHO) based on an engineering, sociotechnical and governance perspective on control aimed at addressing accountability gaps in Autonomous Weapons Systems (AWS). While the focus in that paper is not on accountability perse, Verdiesen et al. (2021) does subscribe to a definition of accountability as a form of control, which aligns with that of Bovens (2007): 'An agent is accountable to a principal if the principal can exercise control over the agent' (Lupia 2003). Furthermore, CHO is an extension to MHC that the authors define as broadening of the concept, which they deem to primarily focus on the "relationship between the human operator and Autonomous Weapon System".

The CHO Framework consists of three horizontal layers that are based on the three-layered model that Van den Berg (2015): (1) technological layer in which the technology is described, (2) the socio-technical layer in which humans and technology interact in activities and (3) the governance layer in which institutions govern these activities. These layers are offset versus developments in three time phases: (1) before deployment of a weapon, (2) during deployment of a weapon and (3) after deployment of a weapon, which describe the environment of the system, which can range from more internal to more external to the technical system. It must be pointed out that Verdiesen et al. (2021) focus their CHO framework solidly on accountability of AWS and hence has some limitation with regard to generic technical autonomous systems. Furthermore, they base their perspective on MHC as defined in Santoni de Sio and van den Hoven (2018), which has been further expanded in recent years by Calvert & Mecacci (2020). Therefore, some of the shortcoming they identify with regard to MHC will be disregarded as they have, at least in part, been addressed in the mentioned literature. Verdiesen et al. (2021) define a notion of narrow Meaningful Human Control, focusing on the operational relationship between one human controller and one technical system, identifying the conditions for effective interaction. This view aligns with that of proximal MHC. On the other hand, distal MHC is coined broad Meaningful Human Control, to consider of autonomous systems that are sufficiently responsive to ethical and societal needs, which was later shown to also include "social institutional and design dimension at a governance level" (Calvert et al. 2024), which Verdiesen et al. (2021) mention as a shortcoming. However, Verdiesen et al. (2021) later state that the broad, or distal, notion of MHC can also be used to fill some gaps that exist in the CHO framework.

In the CHO, the combination of layers and phases result in nine blocks each containing a component of control explicitly focussed on the deployment of AWS. Nevertheless, the idea of defining environment levels of control through a Governance, Socio-technical and Technical layer has relevance when considering other autonomous systems. Moreover, these layers have a clear connection with elements from the MHC tracing condition, aligned with ensuring a chain of control of human agents, demanding that this human has the ability and skill to act, and has a level of moral accountability. The alignment with the MHC proximity scale furthermore allows this approach to be more easily adopted for the further operationalisation of MHC.

**Contestable AI by Design**

On the topic of 'Contestable AI by Design', Alfrink et al. (2022) proposed a design framework following an extensive synthesis of sociotechnical features using qualitative-interpretive methods. Contestable AI by Design is a growing field of research focussed on ensuring that AI systems are responsive to human intervention throughout their system lifecycle, where the contestability equates to the ability for "humans challenging machine predictions" (Hirsch et al. 2017). Within this framework various aspects primarily involving system design for "generic automated decision-making" systems , which resulted in the identification from literature of five features of six practices that the authors claim are a step towards "intermediate-level design knowledge for contestable AI", which are given in Table 1 below.

**Table 1: Principles and practice of contestable AI by design**

| Features | Practices |
|---|---|
| 1. Built-in safeguards against harmful behaviour; | 1. Ex-ante safeguards; |
| 2. Interactive control over automated decisions; | 2. Agonistic approaches to machine learning (ML) development; |
| 3. Explanations of system behaviour; | 3. Quality assurance during development; |
| 4. Human review and intervention requests; and | 4. Quality assurance after deployment; |
| 5. Tools for scrutiny by subjects or third parties. | 5. Risk mitigation strategies; and |
| | 6. Third-party oversight. |

These features and practices are constructed to ensure AI systems are open and responsive to contestation by those people directly or indirectly impacted throughout the system lifecycle and hence protects human self-determination and ensures human control over automated systems throughout the lifecycle of a system (Alfrink et al. 2022). The system requires *built-in safeguards against harmful behaviour*, where a second automated system checks decisions for alignment and flags issues for human review. Shared control is recommended, with final decisions being a negotiation between the system and the user. Users should understand how decisions are made, and decisions must be reproducible and traceable. Human review of system performance to access context and correct harmful decisions is advised as a form of quality control(Almada 2019; Walmsley 2021). Human controllers responding to intervention requests must have the authority and capability to alter previous decisions (Brkan 2019). On a strategic level, there should be processes in place for scrutiny by system users or Third Parties stakeholders, which includes aspects involving documentation and clear Operational Design Domains (ODD) descriptions.

The framework, shown in Figure 2, captures many features and requirements also recommended from other discussed approaches. There is also clear alignment with MHC on various aspects. For example, having human control that is knowledgeable and capable to intervene in a meaningful way on one side, while on the other side is able to make decisions that can be reproducible and traceable, are two main aspects of the traceability condition of MHC. Furthermore, scrutiny and involvement from designers, but also broader stakeholders is a clear demonstration of distal control as defined in Mecacci & de Sio (2019) and allows for explicit distal updating of an autonomous system as defined by Calvert et al. (2024). Therefore, the approach set out by Alfrink et al. (2022)'s framework for contestable AI by Design are agreeable with that of MHC based on tracking and tracing. It should therefore be considered as an important aspect of the process of design of autonomous systems, regardless if they are AI based or not. The context of the framework is limited to design, and specifically to decision-making systems. In their own words, the authors state that the "framework probably does not cover cases… where time-sensitivity of human intervention is relatively low" and that MHC is more suited to such a context. Two further aspects that the framework does not cover is that of explicit system evaluation, as it is a design framework, and aspects relating explicitly to the integration of human reasons into the system aligning with the tracking condition from MHC.

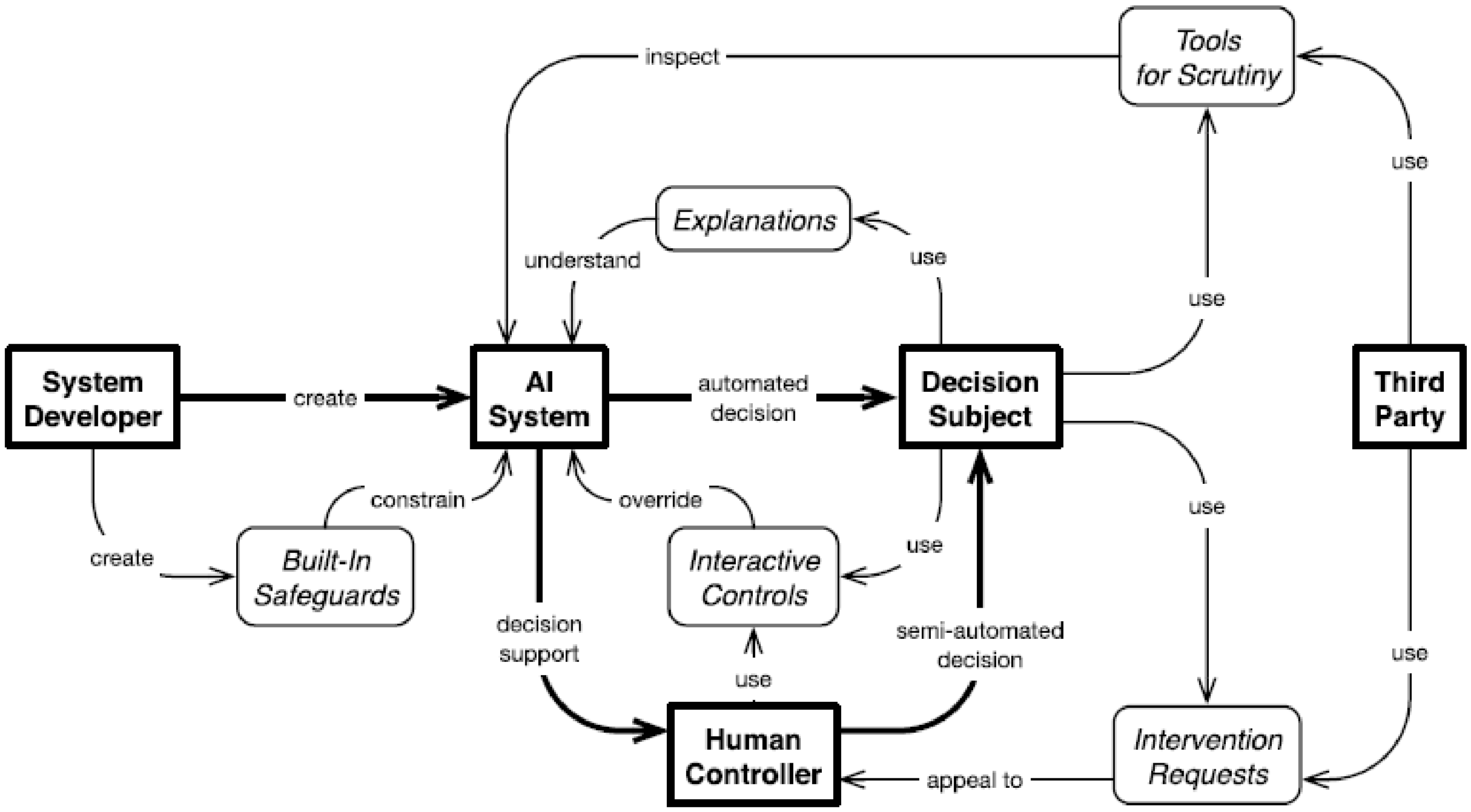

**Figure 2: Overview of features of contestable AI (from Alfrink et al. (2022))**

**Value Sensitive Design**
Finally, Value Sensitive Design (VSD) focusses on the design of technology while accounting for human values, and doing this in a principled and systematic manner throughout the design process (Davis & Nathan 2015; Friedman et al. 2006). VSD in primarily interested in the investigation of values in technology, serving such purposes as stakeholder identification and legitimation, value representation and elicitation, and values analysis (Friedman et al. 2017). At its heart are elements that encourages co-creation and integrated design between different stakeholders through aiding of identification of stakeholders and their values to create alignment, as well as the resolution of potential issues. VSD has value in ensuring distal values and reasons are properly accounted for in design, however it is not directly relate to a system of control and is therefore seen in the context of this research as an enabling set of principles rather than at the heart of questions relating to construction of responsible autonomous control systems.

The review of related concepts to MHC has proven a relevant one that has highlighted important principles that exist with a large degree of alignment for the responsible and meaningful control over autonomous systems. Although some concepts claim to deal with shortcomings of MHC, they appear to actually agree with much of the current state-of-the-art developments of the concept, while other concepts extend MHC, for example to certification or to aspects of liability. A concise summary of the considered concepts, their key elements and connection to MHC is given in Table 2. Again, it must be stressed that this is a limited overview that has been deliberately constrained to recent developments close to the focus area of MHC, as even small divergence into related domains would deviate beyond the scope of this paper and explode the plethora of concepts that can be discussed.

**Table 2: Descriptive summary of concepts related to MHC**

| Concept | Description | Key elements | Described in | Relation to MHC |
|---|---|---|---|---|
| **Meaningful Human Certification (MHCrt)** | Training and certification of autonomous systems and of responsible individuals | Strategic and technological design certification<br>Focus on:<br>- Ethically, Legally, and Safely<br>- 'Better than human' performance<br>- Certified clarity on control responsibility | (Cummings 2019)<br>(Skeete 2021) | Certification of elements relating to the tracing condition of MHC: identifiable human agents with knowledge and ability to act; and tracking: system performance demonstrated. |
| **Trustworthy Autonomous Systems (TAS)** | Principles to ensure trustworthiness of autonomous systems and AI | Reasoning of trustworthiness:<br>- Reliability<br>- Legality<br>Principles:<br>- Monitoring<br>- Taking over<br>- Informing and directing<br>- Understanding (knowledge) | (Johnson et al. 2014; Yazdanpanah et al. 2023; Ekelhof 2019) | Indirect connection to proximal tracing: chain of control to a responsible human agent with knowledge and ability to act. |
| **Comprehensive Human Oversight (CHO)** | Operationalisation of the concepts of accountability, control and oversight based on an engineering, sociotechnical and governance perspective of control | Contextual layers:<br>- Technological layer<br>- Socio-technical layer<br>- Governance layer<br>Deployment time phases:<br>- Before deployment of a weapon,<br>- During deployment of a weapon and<br>- After deployment of a weapon<br>Notions:<br>- Broad and narrow MHC | Verdiesen et al. (2021) | Broad and Narrow MHC align with Distal (operational) and Proximal (strategic / institutional) aspects of MHC. Control and oversight relate to the expanded tracing condition |
| **Contestable AI by Design** | Design principle ensuring AI systems are open and responsive to human intervention throughout their lifecycle | Principles:<br>- Built-in safeguards<br>- Interactive design and control over Automated decisions<br>- Explanations of system behaviours<br>- Human review and intervention requests<br>- Scrutiny by subjects or stakeholders | (Alfrink et al. 2022; Almada 2019; Hirsch et al. 2017; Walmsley 2021; Brkan 2019; Henin & Le Métayer 2021; Lyons et al. 2021; Sarra 2020; Vaccaro et al. 2021) | Various elements of tracing condition captured in principles: identifiable human agents with knowledge and ability to act. Distal aspects of tracing and tracking system design relate to designer and stakeholder involvement. |
| **Value Sensitive Design (VSD)** | Design of technology while accounting for human values | Key focus:<br>- Stakeholder identification and Legitimation<br>- Value representation and elicitation<br>- Values analysis | (Friedman et al. 2017; Friedman et al. 2006; Davis & Nathan 2015) | Indirect connections to values and norms of tracing, and distal aspects of related to design principles. |

# 4 Principles for applying MHC

Meaningful control over autonomous systems has evolved at different rates across domains, depending on criticality and stakeholders' ability and willingness to advance autonomy. In Defence, where MHC originated from discussions on AWS, progress has stalled, leading to other concepts gaining traction. In Transportation, MHC is found where autonomous control of large moving objects is seen as a safety-critical process, with industrial stakeholders in autonomous vehicles demanding a need for more operational descriptions. In Healthcare, the focus is on decision-making, emphasizing joint decision-making between humans and autonomous systems, and exploring when these systems may outperform humans, while the use of RAS is evidence of the first steps towards physical autonomation. Additionally, the analysis of related and connected concepts to MHC have given additional insights into way that MHC can be applied and principles for the operationalisation.

## *4.1 Principles for MHC operationalisation*

From the investigation up to this point, we propose principles applicable for the operationalisation of MHC, based on the current state-of-the-art of MHC and on the considered related concepts and principles set out in cited literature in the previous sections. The principles are derived making use of a conceptual synthesis approach (Andreasen et al. 2015; Jabareen 2009; Nutley et al. 2002) to acquire the key principles most relevant to the operationalisation of MHC. To derive the core principles, the approach emphasizes the systematic decomposition of concepts, iterative refinement, and structured synthesis (Andreasen et al. 2015). Conceptual synthesis requires identifying foundational components, which we have decomposed in Sections 2 and 3 of this paper. The key elements from the concepts (Section 3) and the initial domain-specific frameworks and approaches (Section 2) were analysed and structured using the 'concept integration' method into distinct but interrelated principles based on logical coherence and applicability to autonomous systems, and that are generally agreeable and required to operationalise MHC. These principles are grouped and named in alignment with MHC terminology.

**Distal – Proximal distinction**: also referred to as 'Broad – Narrow', 'Design – Operation' or 'Distributed Control'. *Distal* considers humans (stakeholders) and their reasons at a higher abstraction level with a greater degree of complexity and a longer timeframe (Mecacci & de Sio 2019) (also see Figure 1). At this Governance and Socio-Technical level, society, designers and regulators play a prominent role before autonomous systems are deployed, but explicitly also during their operation to allow adjustment on a strategic design level. *Proximal*, on the other hand, considers agents (both human and machine) close to the operation of an autonomous system on a shorter timeframe after deployment, readily aligned on a Technical level.

**Tracing condition of MHC**: The tracing condition states that there must be one or more human agents in a system's design and operation who are knowledgeable and capable human agents with the ability to act. Moreover, they must appreciate the (in)capabilities of the system, and secondly, understand their own role as targets of potential moral consequences for the system's behaviour. While not the primary focus of this paper, this latter aspect also connects to aspects of responsibility, accountability and liability, as set out for MHC by Santoni de Sio & Mecacci (2021).

**Tracking condition of MHC**: also aligned to the correct functioning of an autonomous system and system responsiveness (as found in Contestable AI). The tracking condition considers the responsiveness of a system's behaviour to human (moral) reasons and intentions to act. This entails that the autonomous system must act in accordance with what is explicitly and implicitly humanly acceptable. Furthermore, the performance of an autonomous system to meet the tracking condition should also be open to contestability to improve and ensure correct functioning, both on a proximal and distal level.

**Integrated system perspective**: MHC is considered over an autonomous system. From the described concepts and literature, it becomes evident that systems can be considered on different interrelated levels and from different perspectives. Therefore, MHC operationalisation must explicitly include consideration of the expanse of the considered system. Moreover, different systems and sub-systems need to be identified and included explicitly, including the interactions between (sub-)systems, such as that of Governance, Socio-technical and Technical systems. Testing of system components is also an essential part of a system perspective.

**Evaluation and Design of autonomous systems**: aligned to backward-looking and forward-looking principles (Van de Poel 2011). MHC should be applied as a concept to govern the design of an autonomous system so that it functions in an acceptable and responsible way, which also includes a sound degree of responsibility and accountability attribution (aligned to forward-looking). This connects with many principles set out in the majority of considered concepts on system design. On the other hand, MHC can also be used to evaluate and monitor already deployed or systems that are being tested, which allows for improvement of system performance.

**Distal and proximal updating**: Improvement and correction of autonomous system design, either on a detailed level or system level, or of human involvement and ability connects the findings of evaluation to on-going design aspects. Distal updating, referred to as explicit or 'by-choice' in Calvert et al. (2024), considers explicit decisions by distal stakeholders (such as designers, regulators, etc) to make or enforce changes to a systems design or a humans role to improve MHC after evaluation. Proximal updating is a more implicit form of system update through a learning process, either by a system able to independently perform self-updates, such as through Machine Learning, or by a human agents in a position of control who has increased their ability to act, for example through training or gaining new insights through experience.

**Co-creation and broad stakeholder involvement**: Both in the design and the evaluation process of an autonomous system under MHC, a broad distal involvement of relevant stakeholders is required. The primary motivations for this lie at encapsulating different elements of human values and reasons, while also including knowledgeable human agents that can have a positive influence of ensuring a greater degree of MHC in the autonomous system.

**Cooperative and joint human-machine control**: also referred to as 'joint human-machine teams', 'symbiotic control' or 'shared/traded control'. Explicit clarity on the roles and responsibilities of an autonomous system and human operator on a proximal level is required. This includes situations in which both have different roles as well as situations in which there is active cooperation and collaboration between humans and autonomous system in system control. One can also state that human intervention is also included here, as well as a clear description of levels of autonomy, which are present in each of the considered domains. This principle therefore overlaps with some other principles, not least that of the tracing condition that also includes the chain of human control aligned with responsibility and accountability. Nevertheless, it is important enough to include separately as it also highlights the type and form of cooperation.

**Ensuring redundancy**: also referred to as built-in safeguards. In safety-critical autonomous processes, allowing system failure can have catastrophic consequences. Redundancy does not currently exist in MHC theory, however the degree of MHC can be increased by its inclusion as found from the above analysis. Inclusion of redundancy can be seen as a design choice, however its inclusion must be considered versus the explicit consideration of what of the consequences are of a lack of MHC, and to what degree this is found to be acceptable.

# 5 Approach for operationalisation of MHC

In this section, we present the framework for the operationalisation of MHC for an autonomous system. This framework is based on the principles that were derived from broad cross-domain literature, as are presented in the previous section, and on the latest state of the art on MHC theory. Firstly, the framework is presented and described in relation to its constituent sub-systems and their significance. This is followed by a justification of how the principles for MHC application are integrated within the framework, along with an explanation of how the MHC conditions are fulfilled. Finally, the section concludes by outlining how the generic framework can be further refined for application in domain-specific contexts.

## *5.1 Generic operationalisation framework for MHC*

The derived framework for MHC operationalisation is shown in Figure 3 and gives an elaboration of potential sub-systems, agents (humans and organisations) and their sphere of control over the autonomous technology from an MHC perspective and in alignment with the derived principles for application of MHC in practice.

The framework exists out of various sub-systems, which, for the description of the elements in the framework, are clustered on three levels based on their proximity to the autonomous technology. The **first** focusses on the **autonomous technology** itself as well as those agents in close proximity to it and able to influence operational control. The **second** considers the first distal layer of agents and processes that can influence the autonomous technology through design and interactive intervention with a focus on **designers, user preparation and environment**. The **third** focusses on the collection of further distal sub-systems and agents relating to regulatory and societal spheres of control, considering **government, regulation and society.** The following paragraphs outline these various generic subsystems, specifying and justifying the various connections and forms of control, which are verified regarding MHC in sub-section thereafter.

**Autonomous Technology**

The first sub-systems to be identified are the key proximal systems that involve the *'system of joint human-machine control'*, which exists of the considered *autonomous technology* together with the technology's *user*, which can be seen at the heart of the framework in Figure 3. The autonomous technology includes the physical mechanical and material elements of the technology, as well as the digital elements, which includes the software and any programmed intelligence. These two 'agents' collectively have operational control over the technological system in differing degrees depending on how the system is designed, setup, and in which modes of automation it operates in its environment. In some cases this joint control may be shared between the agents, while in other time it may be traded, where one merely supervises the other (Abbink et al. 2018).

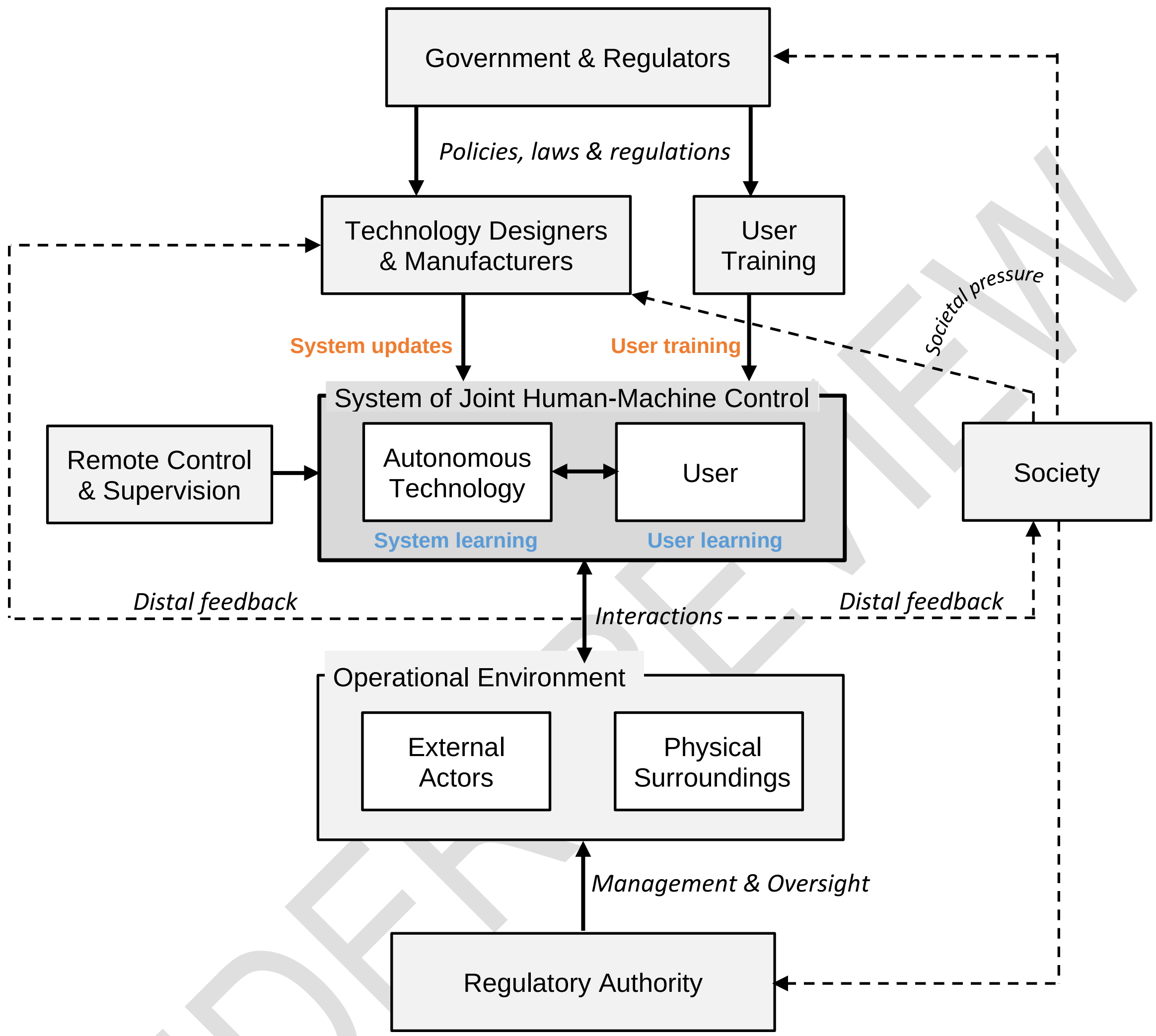

**Figure 3: Generic operationalisation framework for MHC**
**(blue indicates implicit proximal updating, orange indicates explicit distal updating)**

A third agent can be identified on the proximal level in the form of a *remote controller or supervisor*. This agent is on the boundary between proximal and distal influence but has the ability to directly influence proximal behaviour and hence is included on the proximal level. In some systems, this could be a person or external system that oversees correct functioning of the technological system or could in other systems act as a form of redundancy in case the joint human-machine control is in danger of becoming diminished or failing. The autonomous technology and user have the ability to improve operational control, moreover Meaningful Human Control, through learning from experiences on a proximal level, or through receiving updates externally on a distal level. These forms of updates were previously coined *'implicit proximal updating'* and *'explicit distal updating'* respectively (Calvert et al. 2024), and are terms which we will retain here. *Implicit proximal updating* (shown in Figure 3 in blue) occurs with the user primarily as an artefact of increased user experience and greater understanding of their role and the capabilities of the autonomous technology. As the user gains increasing knowledge through experience, they are increasingly capable to exert control when and in forms that are appropriate. Many current and future autonomous technologies, on the other hand, make use of AI and hence have a built-in capability to learn from past events and improve performance, either as an individual agent or part of a collective of autonomous agents. The *explicit distal updating* (shown in Figure 3 in orange) similarly improves the performance of the user and the autonomous technology through increased knowledge and ability, but from an external intervention. For the user, this takes on some sort of explicit training, while the autonomous technology may receive system updates enabling it to exert increased control.

**Designers, User Preparation and Environment**

The first layer of distal systems are that directly influence the system of joint human-machine control, is the *technology designers and manufacturers*. These are the agents and processes that involve technological system design and production. In some cases, it may be desirable to separate these two elements into two separate sub-systems depending on the way they operate in practice. This sub-system has an initial influence on how the technical system is designed, but also in most cases plays a role in the continuous updating and improvement of the technological system, such as through explicit distal updating for software updates that may occur. On a similar level and focussed on a proximal human agent is the sub-system of *user training*, which has a direct influence on the knowledge of the user or operator of the technological system. This is also a sub-system that has the ability to externally update and improve the performance of the joint human-machine control system through continuous or periodical education or training of the user.

Conversely, the immediate environment in which the autonomous technology is active is a key sub-system in which direct interaction with the autonomous technology occurs. This *operational environment* exists out of potential *external agents* as well as the *physical environment*. External agents can be humans or other technical systems within the immediate environment who may directly interact with the core system or are indifferent to it, but can still affect its operation. The physical surroundings are generally static elements that define physical constraints of the autonomous technology's movement or area of influence. Interactions between agents on this level will generally not lead to explicit updating of the technology, but can lead to proximal updating within the joint human-machine control system through learning from new or recurring experiences, which can be present either or both with the autonomous technology and the user.

**Government, Regulation and Society**

On an even greater distal level, there are sub-systems and agents that do not have an as direct influence on the human-machine control system, but can indirectly exert distal control on those sub-systems that do directly influence that sub-system. In the first place, *Government and Regulators* play a key role in dictating the boundary conditions and constraints of how the autonomous technology can be used and which requirements are necessary in the design of it. This can take the form of policies, laws or regulations that are set out and can be enforced by a *regulatory authority* with oversight and sometimes also management of the environment in which the autonomous technology interacts. Similarly, these authorities (both legislature and overseeing) can also dictate and oversee other proximal sub-systems including that of the technology user or remote controller. Beyond government and regulation, *society* is the most distal sub-system considered and plays a fundamental role in determining and influencing some of the key aspects of what MHC entails and what is deemed acceptable on a societal level. Society, as a sub-system, is complex and we explicitly do not delve deep into those complex dynamics that influence the autonomous technology in this paper. However, on a more general level, society is where many norms and values are held in an expansive, heterogenous and dynamic way. Societal pressure can be exerted on government and regulators, as well as on vehicle designers or even human users of autonomous technology, which in turn can force explicit distal updating of the autonomous technology or affect the role of the user in the human-machine control system. Often, a trigger is required for societal pressure to occur, which can often be the result of the performance of the human-machine control system in its interaction with and in the operational environment. This is depicted in the generic framework through the distal feedback from these interactions shown in Figure 3.

## 5.2 Application of MHC principles

The principles for MHC operationalisation are at the core of the presented framework. The inclusion of the principles is achieved through first grouping the principles into five different categories that can be collectively applied to the operationalisation of MHC. The first step starts with *explicit sub-system identification* for the framework. These sub-systems can and should be identified explicitly as either proximal or distal, aligning with their proximity to the physical operational control of the autonomous system. This includes elements of governance (distal), socio-technical (can be distal or proximal) and technical (primarily proximal). The further four categories of principles are captured in *proximal process*, *distal process*, elaboration of the *tracing condition*, and of the *tracking condition*, as shown in Figure 1.

Within the *proximal process*, agents (both human and autonomous) are identified within proximal sub-systems that have the ability to operationally control the technical system. Each agent is evaluated for their ability to exert control and to also obtain proximal updating that would further enhance their knowledge and ability to exert control. The proximal control process should also be disaggregated in a separate overview to the complete system framework.

On a *distal process level*, we start with the identification of agents and stakeholders who can distally exert control or influence over the technical system, either during the design, evaluation or operational phase, as these are agents that connect both to the proximal sub-systems (autonomous technology and user) and to the more external agents. On the first distal level close to the autonomous technology, these can be agents that directly design the system, while many will be organisations and sub-systems that have a less pronounced impact on the systems design, such as on a governance level for example. Also, within this process, the ability of each agent to influence the design and control process is investigated as well as their ability to create feasible and realistic possibilities for explicit distal updating. An important aspect of this category is explicit consideration of options for system evaluation during testing and operation.

The *tracing operationalisation* involves identifying the chain-of-control from agents to the technical system. How can and do agents exert control and to what extent should this full or shared control be elaborated? Assessment or fail-safing of agent's ability and knowledge forms part of this category and can be extended with formal or informal certification (e.g. training or licencing). Therefore, options of redundancy are included in this category. System design specification will determine if and level of redundancy is required, while the options for redundancy can be identified through the available agents and their chain-of-control and abilities. Finally, proximal updating can be checked and adjusted or expanded during this phase.

The *tracking operationalisation* focusses on the autonomous systems ability to meet with human intentions and reasons. These are aspects that lie deep in the AI and autonomous system technical design, which lies outside the scope of this paper. Nevertheless, inclusion of where these aspects are included are included in the approach on a higher abstraction level. Design of autonomous AI systems is a process that should include co-creation principles, and hence direct lines of involvement from distal agents can be detailed in the design and the continuous evaluation process, even after implementation and operation. An explicit part of this also connects to identifying and further improving the options or explicit distal updating, as well as identification of redundancy options for the technical autonomous system, which could even include AI systems monitoring other AI systems, as suggested in Alfrink et al. (2022).

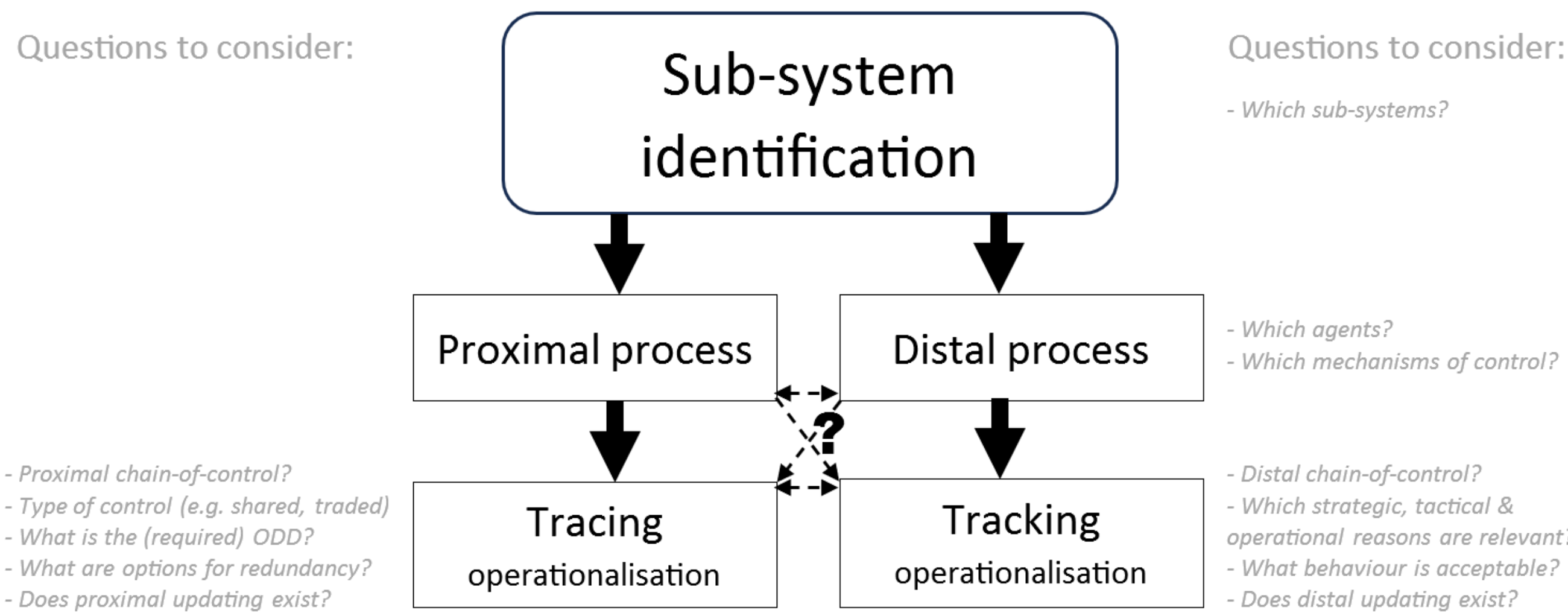

**Figure 4: Process to apply MHC principles to construct operational diagram for an arbitrary autonomous system. Including prompting questions for assistance.**

The five categories that make up the process of constructing an operationalisation diagram based on the principles for MHC operationalisation are highlighted in Figure 4 with some accompanying questions to aid the process of diagram construction and are based on the described categorise as given above and descriptions from the previous section.

## *5.3 Verification of Tracing and Tracking and principles*

The complete generic framework is constructed from an integrated system perspective, in which generic sub-systems for autonomous technology operation are identified and given their place in operations. Both proximal and distal sub-systems are identified as well as their ability to lead to a learning process through either proximal or distal updating of the joint human-machine control system to improve MHC. Details of the form and extent of cooperative or joint human-machine control are too application specific to be explicitly generalised and require detailing depending on the considered autonomous system. The same holds for the design principles of the autonomous system, which in itself is a complete field of research that we do not explicitly dive into in this paper. Nevertheless, the design process should include elements of co-creation, which also includes different stakeholders, some of which are mentioned in the generic framework in the form of government, regulators and society. Ensuring redundancy in a system is included to some extent through the possibility for remote control or supervision, as well as building redundancy into various elements of the autonomous technology and having the user act as a form of redundancy. Each autonomous technology will have different requirements in this regard and there will be differing levels of legislative requirements for redundancy. Finally, the tracing condition is explicitly captured in the framework through creating a clear chain of control through different agents in the sub-systems, and through identification of human user and agents that can exert control and for which a clear awareness and ability must be present. The tracking condition is more difficult to capture in the framework, which entails that the autonomous systems must align and act with human reasons. These reasons will come from various human agents, both proximal and distal as shown in the framework, but need to be explicitly defined and validated for an arbitrary system. Regardless, the framework lays the foundation for this to be easily conducted in practice.

## *5.4 Domain-specific contextualisation*

While the focus in the paper is on a framework that gives a generic depiction for an arbitrary autonomous technology, the framework can easily be extended and made more explicit and applicable for domain-specific autonomous systems. Some thoughts on how this can be performed are given in this sub-section. While the focus of this paper is on a generic framework for Meaningful Human Control (MHC) for autonomous systems, the framework can be adapted for specific domains by considering their unique requirements, risks, and regulatory landscapes. This subsection offers some thoughts and an initial approach to domain-specific contextualization.

Starting with the presented generic framework, the first step in this process is *the alignment with the generic framework*, which involves *identifying domain-specific constraints and requirements*, particularly those shaped by ethical concerns, safety-critical operations, and legal mandates and aligning these with the generic framework. In additional, the *MHC principles should be mapped* onto the operational processes within the considered domain. This involves analysing how responsibility attribution, oversight mechanisms, and contestability structures can be adjusted to sector-specific needs. Secondly, *decision making and domain-specific risk and mitigation strategies* should be defined and integrated where applicable. This requires defining clear intervention points where human control and intervention is necessary or desirable, such as escalation protocols for critical decisions or fail-safe mechanisms for manual override in emergency scenarios. A final suggested step is *validation through stakeholder engagement*. It is crucial to engage stakeholders, regulatory bodies, and industry practitioners in an iterative validation process. By involving stakeholders, the framework can be refined through real-world applications and policy discussions. This iterative feedback loop ensures that the adapted framework remains practical, contextually relevant, and compliant with evolving regulations, and ensures that MHC can become embedded in the system ensuring responsible control.

In summary the suggested steps are:

1. Alignment with the generic framework and MHC principles
    a. Identification of domain-specific constraints and requirements
    b. Mapping of MHC principles
2. Integration of domain-specific decision making and risk and mitigation strategies
3. Validation through stakeholder engagement

To illustrate in an example, we take the healthcare domain.  Applying the proposed approach, the first step is to align the domain-specific constraints with the generic MHC framework. Healthcare AI systems operate under strict ethical and legal requirements, including patient autonomy, informed consent, liability concerns, and regulatory standards such as GDPR and medical device approval laws. The generic MHC principles must be mapped onto these requirements to ensure AI-driven diagnoses remain transparent, accountable, and aligned with medical ethics. For instance, oversight mechanisms must explicitly define whether physicians, hospitals, or system providers hold ultimate responsibility for AI-generated diagnoses.

Next, the integration of domain-specific decision-making processes and risk mitigation strategies is necessary to ensure safe AI deployment in clinical settings. This involves defining intervention points, such as requiring physicians to review and approve AI-generated diagnoses before treatment decisions are made. Fail-safe mechanisms, such as alerting clinicians when AI confidence is low or when anomalies are detected, further ensure that human judgment remains central to decision-making. Risk mitigation strategies should also address bias detection and explainability, ensuring that AI systems trained on specific population datasets do not introduce systemic biases that could lead to misdiagnoses in underrepresented groups. Finally, validation through stakeholder engagement ensures the adapted framework remains practical and effective in clinical settings. This involves collaboration with doctors, hospital administrators, regulatory bodies, and patient advocacy groups, allowing the framework to be refined based on real-world applications and evolving medical policies. Embedding MHC principles allows the framework to be adapted to healthcare AI systems, supporting human decision-making and ensuring meaningful human control.

# 6 Discussion

The presented generic framework for the operationalisation of MHC offers an invaluable resource for researchers, industry and government working with autonomous systems to achieve responsible control. It helps identify stakeholders with varying degrees of control, highlights design gaps related to MHC, and supports evaluating a system's ability to perform tasks responsibly. Each sub-system can be separately elaborated in greater detail according to MHC principles, which can assist in (re)designing and validating the sub-systems, and regulations that may be in place, and could lead to adjustments in these processes. The approach and presented framework do come with limitations that we also address in this section considering the validity of MHC as a concept and the way the framework can be applied and used for design and evaluation, followed by considerations of MHC in the broader context of responsible control. We first give some thoughts on MHC's potential application in regulation and alignment with current legislation.

**Legal alignment and positioning on MHC**

At the time of writing, the European Union's AI Act (European Union 2024) is currently the most elaborate piece of regulation, which stipulates regulations for the development, deployment, and use of AI systems. The principles of MHC align closely with the objectives of this AI Act, particularly in ensuring human oversight, accountability, and risk mitigation in AI systems. The AI Act introduces binding legal requirements for high-risk AI systems, mandating human-in-the-loop mechanisms to prevent undue harm to fundamental rights, safety, and democratic values. MHC expands on these legal safeguards by advocating for a proactive and context-sensitive approach to human responsibility in AI decision-making that extends beyond high-risk classifications. While the AI Act provides a regulatory foundation, MHC principles emphasize the operationalisation of human agency in AI governance, to ensure that systems remain ethically contestable and value-aligned. This suggests that MHC can serve as a guiding framework for refining regulatory standards, helping policymakers address emerging ethical and accountability challenges as AI technologies evolve.

Beyond the EU AI Act, MHC also aligns with the likes of the Digital Services Act (DSA) (European Commission 2022) and the proposed AI Liability Directive (European Union 2022), which emphasize transparency, accountability, and redress mechanisms that are inherent to MHC. Moreover, the Ethics Guidelines for Trustworthy AI (European Commission 2019) issued by the EU's High-Level Expert Group on AI promote human oversight as a core requirement for responsible AI development, further reinforcing MHC's relevance in regulatory discourse. By integrating MHC principles into existing and emerging legislation, regulators can enhance legal certainty around AI accountability while ensuring that human control mechanisms are contextually appropriate and enforceable. The alignment between MHC and current legislative efforts underscores MHC's potential as a normative and operational framework, which can help guide AI governance in ways that uphold both ethical responsibility and legal compliance.

**Final thoughts on validity of MHC**

Since its inception in 2015, MHC has been seen as a highly promising concept to comprehensively understand and deal with responsible control over autonomous systems despite some setbacks. In the meantime, several related concepts have emerged that closely resemble MHC, as discussed in Section 3. Aligning with these concepts has helped strengthen the foundation for MHC's operationalization and application. We acknowledge that some scholars may disagree or take a more nuanced approach, which is valid. However, we argue that the foundations of MHC are now firmly established. Its broad, encompassing nature, once seen as a challenge for operationalization, is actually a key strength, allowing for a comprehensive approach to controlling autonomous systems, unlike other narrower concepts. By integrating developments from these related concepts into the MHC framework, greater progress can be achieved. While some may reject the premise of MHC, we and others have demonstrated its promise and utility—possibly even its necessity—in controlling autonomous systems. The often-cited stumbling block of clarity and applicability is addressed in this work, offering clearer pathways for operationalization across different domains.

**From strategic design to assessment for MHC**

The framework aids in the strategic design and evaluation of autonomous systems, but will require further detailing per sub-system, which cannot be done effectively on a generic level. Different domains have distinct regulations, processes, and contexts regarding criticality, which must be considered when applying the framework. After adapting the framework for a particular autonomous technology, it may be desirable to assess the extent to which MHC is present in an existing system's design. Currently, research is on-going that is aimed at qualitative and quantitative evaluation of MHC for responsible control. Qualitatively, the previously discussed conditions for MHC (Santoni de Sio & Van den Hoven 2018) and derived principles in this paper are a good starting point for the qualitative evaluation of MHC. Research on the quantification of MHC for assessment is also on-going, although naturally appears to be very domain specific with regard to applications.

**Considerations of joint human-machine control in the context of shared and traded control**
The framework for joint human-machine control includes both shared and traded control, as outlined by Abbink et al. (2018). While this paper does not delve into robot autonomy and control, which has been extensively covered (Abbink et al. 2018; Kim et al. 2024; Onnasch & Roesler 2021), it is important to note the connections between this work and the hierarchical framework for shared control in Abbink et al. (2018). That framework defines strategic, tactical, operational, and execution task levels where control can be shared or traded between human and robot. It also considers knowledge-based, rule-based, and skill-based interactions. Further research exploring joint human-machine control from an MHC perspective is one that we recommend as an interesting if not essential follow-up piece of research that could give greater depth to the joint human-machine part of the framework from an MHC perspective combined with the shared-traded control given in the cited research.

**Final remarks**
To conclude, it should be highlighted that this framework, just like any other approach or model, should be considered as guidance rather than a process to be blindly followed. A model is a representation of reality, likewise, the framework represents a strategic overview of control processes that, in practice, can often be more complex and nuanced than in a framework, which should be considered when applying the framework. Furthermore, we emphasise that this is given as a generic framework that can be used as a starting point to fill in domain-specific characteristics for application for any autonomous system. This final point is one that, while this framework aims for generality, the operational context, system specificity and knowledge of a specific domain must also be incorporated when a framework or concept is applied in practice (Ekelhof 2019).

# 7 Conclusions

Autonomous systems are becoming integral to society, with expanding applications and increasing autonomy and complexity. Many of these systems interact closely with humans and hence require to perform in a responsible, accountable and transparent way. Meaningful Human Control (MHC) is viewed as a concept that enables these aspects to be catered for in the design and evaluation of these systems. However, making such a philosophical concept readily applicable has proven difficult. This paper has set out the principles for the operationalisation of MHC over autonomous systems and proposed a generic framework for the operationalisation of MHC, allowing the concept to be applied to various critical autonomous systems. The framework includes actors close to the technology that exert direct (or proximal) control, as well as stakeholders, such as system designers, regulatory and society, which can influence the performance of the system more distally, such as through software-updates, policy, regulation and societal pressure. When applied to specific domains and autonomous systems, the framework acts a foundation to which more explicit domain-specific detailing can be added to ensure and increase the degree of MHC that autonomous systems can exert and in turn increase responsibility, accountability and transparency, as well as system safety. It is also the hope of the author that the framework will increase clarity on the necessity of using MHC for autonomous systems in industry and public organisations and will break the impasse of applying a highly relevant and necessary philosophical concept to technical systems.

# Acknowledgements

<removed for the review process>